\pdfoutput=1
\documentclass[10pt, conference,]{IEEEtran}
\IEEEoverridecommandlockouts
\usepackage{cite}
\usepackage{amsmath,amssymb,amsfonts}
\usepackage{algorithmic}
\usepackage{graphicx}
\usepackage{textcomp}
\usepackage{makecell}
\usepackage{booktabs}
\usepackage[table,dvipsnames]{xcolor}
\usepackage{subfigure}
\usepackage[ruled,linesnumbered, vlined]{algorithm2e}
\usepackage{balance}
\usepackage{enumitem}
\usepackage{multirow} 
\usepackage{ulem}
\usepackage{amssymb}
\usepackage{bbding}
\usepackage{pifont}
\usepackage{tikz}
\newcommand*\circledwhite[1]{\tikz[baseline=(char.base)]{
            \node[shape=circle,draw,black,inner sep=0.5pt] (char) {#1};}}

\definecolor{CellFst}{HTML}{FF9999}
\definecolor{CellSnd}{HTML}{FFCC99}
\definecolor{CellTrd}{HTML}{FFFFFF}

\definecolor{redbrown}{RGB}{165, 42, 42}

\def\BibTeX{{\rm B\kern-.05em{\sc i\kern-.025em b}\kern-.08em
    T\kern-.1667em\lower.7ex\hbox{E}\kern-.125emX}}

\begin{document}

\title{Jupiter: Fast and Resource-Efficient Collaborative Inference of Generative LLMs on Edge Devices}

\author{
    \IEEEauthorblockN{Shengyuan Ye$^\blacklozenge$, Bei Ouyang$^\blacklozenge$, Liekang Zeng$^\lozenge$, Tianyi Qian$^\blacklozenge$, Xiaowen Chu$^\blacktriangle$, Jian Tang$^\vartriangle$, Xu Chen$^\blacklozenge$$^*$
    }
    \IEEEauthorblockA{$^\blacklozenge$School of Computer Science and Engineering, Sun Yat-sen University, Guangzhou, China}
    \IEEEauthorblockA{$^\lozenge$The Chinese University of Hong Kong, Hong Kong SAR, China}
    \IEEEauthorblockA{$^\blacktriangle$Data Science and Analytics Thrust, HKUST (Guangzhou), Guangzhou, China}
    \IEEEauthorblockA{$^\vartriangle$Midea Group, China}
    \IEEEauthorblockA{\{yeshy8, ouyb9, qianty\}@mail2.sysu.edu.cn, zenglk3@gmail.com}
    \IEEEauthorblockA{xwchu@ust.hk, tangjian22@midea.com, chenxu35@mail.sysu.edu.cn}
    \thanks{$^*$Corresponding authors.}
}

\maketitle

\begin{abstract}
Generative large language models (LLMs) have garnered significant attention due to their exceptional capabilities in various AI tasks. Traditionally deployed in cloud datacenters, LLMs are now increasingly moving towards more accessible edge platforms to protect sensitive user data and ensure privacy preservation. The limited computational resources of individual edge devices, however, can result in excessively prolonged inference latency and overwhelmed memory usage. While existing research has explored collaborative edge computing to break the resource wall of individual devices, these solutions yet suffer from massive communication overhead and under-utilization of edge resources.
Furthermore, they focus exclusively on optimizing the prefill phase, neglecting the crucial autoregressive decoding phase for generative LLMs.
To address that, we propose \texttt{Jupiter}, a fast, scalable, and resource-efficient collaborative edge AI system for generative LLM inference.
\texttt{Jupiter} introduces a flexible pipelined architecture as a principle and differentiates its system design according to the differentiated characteristics of the prefill and decoding phases.
For prefill phase, \texttt{Jupiter} submits a novel intra-sequence pipeline parallelism and develops a meticulous parallelism planning strategy to maximize resource efficiency; For decoding, \texttt{Jupiter} devises an effective outline-based pipeline parallel decoding mechanism combined with speculative decoding, which further magnifies inference acceleration.
Extensive evaluation based on realistic implementation demonstrates that \texttt{Jupiter} remarkably outperforms state-of-the-art approaches under various edge environment setups, achieving up to $26.1\times$ end-to-end latency reduction while rendering on-par generation quality. 
\end{abstract}

\section{Introduction}
The emergence of generative large language models (LLMs) has attracted widespread attention from both industry and academia owing to their exceptional capabilities in a wide range of artificial intelligence (AI) tasks. These models, widely deployed in cloud datacenters equipped with powerful server-grade GPUs, have driven increasing intelligent edge applications such as ChatBot \cite{chatgpt2024} and smart-home AI agent \cite{king2024sasha}.
While born on datacenter warehouse, there is an emerging trend of serving LLMs on more accessible edge platforms rather than uploading requests to remote clouds owned by commercial companies, owing to the sensitive and privacy-critical nature of user data. 
A recent survey \cite{li2024personal} on LLM-based edge applications has revealed that over 80\% of industry experts believe personal LLMs should be fully or primarily hosted at the edge to ensure privacy-preserving model inference.

\begin{figure}[t!]
    \setlength{\abovecaptionskip}{-0.1cm}
    \centering
    \includegraphics[width=0.87\linewidth]{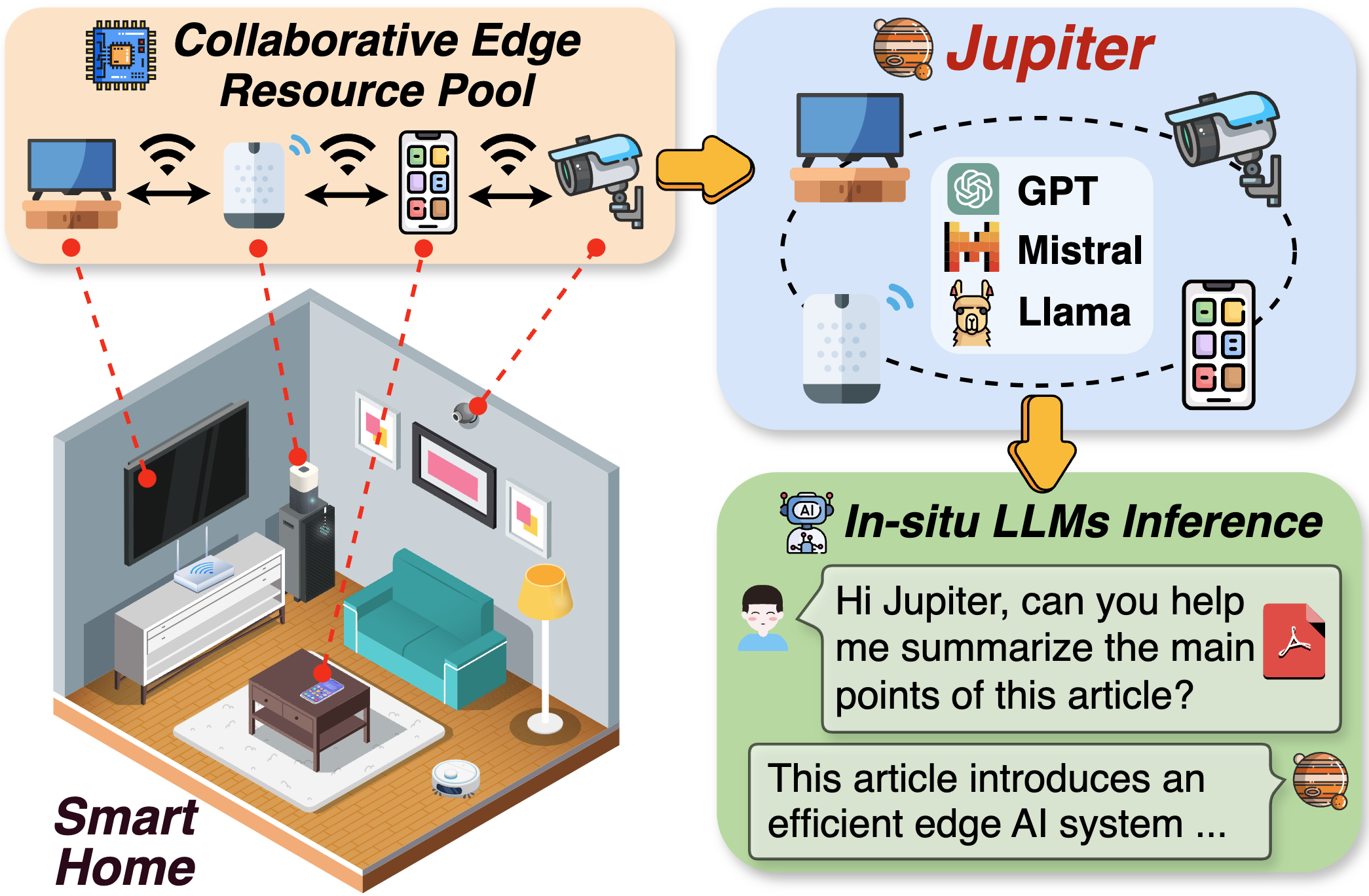}
    \caption{Collaborative LLMs inference in smart home empowered by Jupiter.}
    \label{fig:intro}
    \vspace{-15pt}
\end{figure}

However, hosting computation-intensive and resource-hungry LLMs at the edge is significantly challenging due to the limited resources of individual edge devices, leading to prolonged inference latency and overwhelmed memory footprint.
To address this, some pioneering research \cite{ye2022eco, ye2024asteroid, zeng2024implementation, zeng2025edge} alternatively observe that common edge environments, such as smart homes, usually comprise a variety of trusted idle devices in physical proximity (e.g., phones, laptops, and smart-home devices) that are often connected to the same local area network (LAN). This motivates us to consider them as a collaborative edge resource pool to facilitate in-situ expedited and resource-efficient LLM inference, as depicted in Fig. \ref{fig:intro}.

Motivated by above insight, researchers have proposed different ways for LLM inference with collaborative edge computing \cite{ye2024galaxy, wei2024communication, zhang2024edgeshard}. 
However, these approaches suffer from significant limitations. 
In contrast to datacenter deployment, in-situ LLM serving for edge applications mainly focuses on low latency in processing a single input sequence (e.g., for the purpose of context/intention-aware intelligent control or response in smart homes).
Galaxy and its subsequent works \cite{ye2024galaxy, wei2024communication} borrow ideas from tensor parallelism \cite{narayanan2021efficient} to achieve parallel acceleration of single-sequence inference across multiple edge devices. 
However, these methods necessitate multiple tensor synchronizations at each decoder layer to ensure inference correctness, making communication latency a bottleneck under low-bandwidth edge environments.
Other few works, such as EdgeShard \cite{zhang2024edgeshard}, leverage pipelined architecture \cite{huang2019gpipe} to orchestrate collaborative edge devices. 
However, in single-sequence request scenarios, it ultimately degrades to sequential inference, failing to leverage the computational resources of multiple edge devices concurrently.
Furthermore, all the aforementioned works have focused exclusively on optimizing the prefill phase, neglecting the decoding phase, which is also crucial for generative LLMs.

In this paper, we address the limitations of existing systems by introducing \texttt{Jupiter}, a fast, scalable, and resource-efficient collaborative edge AI system for generative LLM inference, guided by the following design goals: 
(1) Enable parallel acceleration of single-sequence inference during both prefill and decoding phases. (2) Reduce memory footprint per device by distributing LLM parameters across participating devices. (3) Minimize tensor exchanges to ensure robust inference performance in low-bandwidth edge environments.

To achieve aforementioned design goals, we eschew tensor parallelism but instead adopt a pipelined architecture as a principle to orchestrate collaborative edge devices. This architecture assembles the memory of multiple edge devices to support the target LLM while maintaining high communication efficiency, as devices exchange only a small set of activations with their neighbors. 
To enable resource-efficient pipeline parallel inference for single-sequence requests, we leverage the key property of generative LLMs and propose a novel intra-sequence pipeline parallelism method. To maximize resource utilization, we introduce a novel dynamic programming-based parallelism planning for optimal LLM and sequence partitioning, taking into account device heterogeneity, memory budget, and varying input lengths. 
To conquer the parallelization of the autoregressive decoding phase, we first incorporate the idea of speculative decoding into our collaborative inference system to enhance resource efficiency. 
Next, to further boost parallelism potential by leveraging multiple edge devices concurrently, we borrow the wisdom from human thinking and further introduce an outline-based pipeline parallel decoding method.
Extensive evaluations on practical edge testbeds demonstrate that \texttt{Jupiter} achieves up to $26.1\times$ end-to-end latency reduction compared to baselines while maintaining significant scalability in bandwidth-limited environments.

In summary, this paper makes the following contributions.
\begin{itemize}[leftmargin=*]
    \item Through extensive measurement studies on existing edge collaborative LLM inference systems, we advocate a pipelined architecture as a principle to orchestrate edge devices for fast generative LLM inference.
    \item We address the challenge of pipelined parallel acceleration during the prefill phase by proposing an intra-sequence pipeline parallelism method, combined with meticulous parallelism planning to maximize resource efficiency.
    \item We achieve parallel acceleration of the autoregressive decoding phase by integrating speculative decoding into our collaborative inference system and introducing an outline-based pipeline parallel decoding method that efficiently utilizes multiple edge devices concurrently.
    \item We implement \texttt{Jupiter} and evaluate it in realistic edge testbeds. Experimental results show up to $26.1\times$ latency reduction over the state-of-the-art methods, while maintaining significant scalability in bandwidth-limited environments.
\end{itemize} 

\section{MOTIVATION AND PRELIMINARIES}

\subsection{Decoder-Based Generative LLMs Architecture}
Decoder-based LLMs, such as GPT-3\cite{brown2020language}, LLaMa\cite{touvron2023llama}, and Mistral\cite{jiang2023mistral}, have revolutionized natural language processing by enabling tasks ranging from simple text generation to complex problem-solving and conversational AI. In this paper, we focus on deploying these powerful generative models on collaborative edge devices.

\begin{figure}[t!]
    \setlength{\abovecaptionskip}{-0.1cm}
    \centering
    \includegraphics[width=0.95\linewidth]{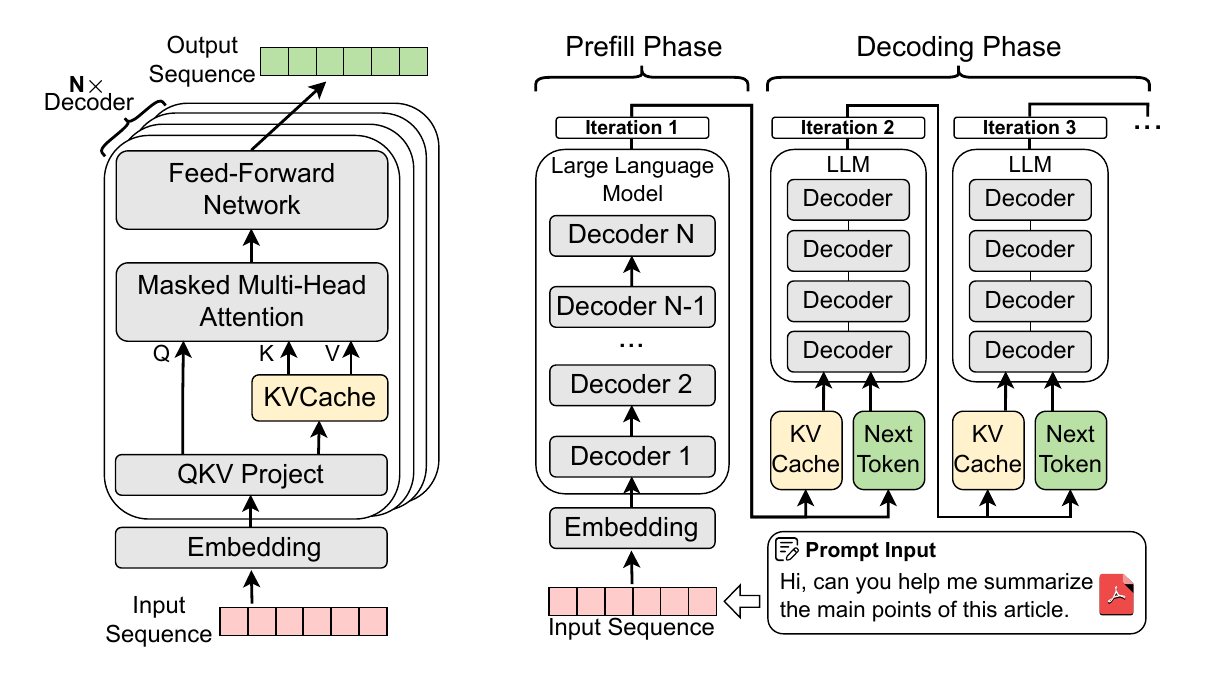}
    \caption{Left: The architecture of a decoder-based LLM. Right: An instance of prefill and autoregressive decoding phases during generative LLMs inference.}
    \label{fig:background}
    \vspace{-15pt}
\end{figure}

As shown in Fig. \ref{fig:background}(Left), a typical decoder-based LLM comprises input embeddings and multiple sequentially stacked decoder layers. Each decoder layer contains several key modules: (1) \textit{QKV Project} takes the input tokens and transforms them into three distinct representations: queries (Q), keys (K), and values (V). (2) \textit{Masked Multi-Head Attention (MHA)} performs self-attention for each head independently, concatenates their outputs, and processes them through a final linear layer. Masking ensures each position only attends to previous positions, maintaining the autoregressive property. (3) \textit{Feed-Forward Network (FFN)} involves two linear operations that first expand the hidden size to a larger dimension and then compress it back to its original size. (4) \textit{KVCache} serves as a dynamic repository where the keys and values of all the previous tokens are typically memoized, allowing models to access and reuse previously computed information expediently.

\subsection{Generative LLMs Inference and KV Caching}
Text generation with LLMs involves two main phases:
\subsubsection{Prefill Phase} 
The initial delay after submitting a prompt input (i.e., time-to-first-token) is the processing time during the prefill phase, which occurs only once per input sequence. As illustrated in Fig. \ref{fig:background}(Right), the LLM first takes the prompt sequence input and predicts a new token that serves as the initial token for the decoding phase. The intermediate states, including keys and values for each decoder layer, are stored in the \textit{KVCache} for reuse in subsequent iterations. 

Existing studies \cite{beltagy2020longformer, li2023sequence} indicate that longer prompts often improve response quality and coherence, driving the ongoing effort to build LLMs that can accept increasingly longer inputs. However, long contexts pose a challenge to response-generation latency during the prefill phase, since the computational load for processing long contexts grows super-linearly with context length due to self-attention mechanism.

\subsubsection{Autoregressive Decoding Phase}
As shown in Fig. \ref{fig:background}(Right), the newly generated token from prefill phase is then fed back into the decoding phase as input, creating an autoregressive process for token generation. To generate a new token that aligns with the context, LLMs must compute its relationship with all previous tokens. The \textit{KVCache} stores the previously computed keys and values of these tokens, enabling their direct reuse without recomputation in each iteration.
Decoding phase is repeated until a stop token is generated or the maximum sequence length is reached.

For tasks requiring the generation of numerous tokens, the decoding phase can dominate the inference process. The challenge with the decoding phase is its autoregressive nature, which makes it challenging to run in parallel and therefore hard to accelerate with multiple edge devices.

\subsection{Collaborative Edge Computing for Generative LLMs} 
\begin{figure}[t!]
    \setlength{\abovecaptionskip}{-0.1cm}
    \centering
    \includegraphics[width=0.95\linewidth]{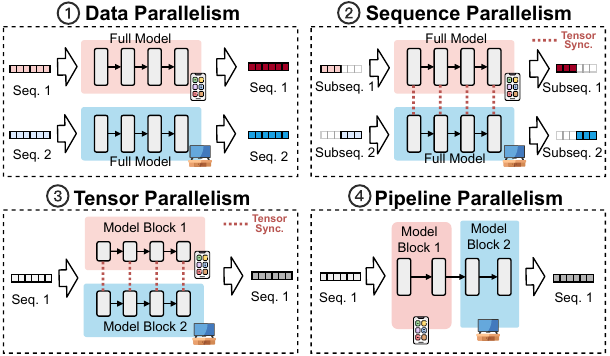}
    \caption{Different parallel inference methods for LLMs.}
    \label{fig:paral}
    \vspace{-15pt}
\end{figure}

\subsubsection{Parallel Inference Methods for LLMs} 
To fully harness the potential of collaborative edge devices for serving generative LLMs, the key question is the choice of parallelism method. We illustrate different parallelism strategies in Fig. \ref{fig:paral}. Specifically, \circledwhite{1} \textit{Data Parallelism (DP)} partitions workloads across the sample dimension, with each device maintaining a full replica of the model and independently performing inferences. However, for single-sequence requests, DP fails to leverage multiple edge devices concurrently. \circledwhite{2} \textit{Sequence Parallelism (SP)} requires each device to maintain a full model replica and partitions the input sequence into sub-sequences for parallel operation, but it necessitates two all-gather synchronizations per decoder layer to ensure consistent inference results. \circledwhite{3} \textit{Tensor Parallelism (TP)} partitions LLM weights across devices, with each hosting a subset of parameters. However, it requires two all-reduce synchronizations per decoder, one after the MHA module and another after the FFN module. \circledwhite{4} \textit{Pipeline Parallelism (PP)} horizontally partitions the LLM into consecutive stages along the layer dimension, with each stage mapped to a distinct edge device. Multiple input sequences are injected into the pipeline concurrently to increase parallelism. However, facilitating parallel inference for a single sequence remains challenging.

We analyze and summarize the model memory usage, computation latency, and total communication volume for various parallelism methods in single sequence inference, as detailed in Table \ref{tab:complexity}. $P$ represents the total number of LLM parameters, $C$ denotes the total floating-point operations for single sequence inference, $L$ indicates the number of decoder layers, $S$ stands for the input sequence length, and $H$ is the size of the hidden state of LLM. We can observe that \textit{PP} is far more communication-efficient than \textit{SP} and \textit{TP} ($L \gg N$) because each device only needs to exchange a subset of output activations with neighboring workers, eliminating the need for multiple tensor synchronizations at each layer.

\begin{table}[t!]\setlength{\tabcolsep}{5.5pt}
\centering
\setlength{\abovecaptionskip}{0cm}
\setlength{\belowcaptionskip}{-0.1cm}
\caption{Analysis on complexity of various parallelism methods.}
\label{tab:complexity}
\begin{tabular}{c|c|c|c|cc}
\toprule[1pt]
\textbf{Features}                                                          & \circledwhite{1} \textbf{DP} & \circledwhite{2} \textbf{SP} & \circledwhite{3} \textbf{TP} & \multicolumn{1}{c|}{\circledwhite{4} \textbf{PP}} & \textbf{\textcolor{redbrown}{Jupiter}} \\ \hline \hline
\begin{tabular}[c]{@{}c@{}}Model Memory\\ Usage (per device)\end{tabular} & $\mathcal{O}(P)$             & $\mathcal{O}(P)$             & $\mathcal{O}(\frac{P}{N})$   & \multicolumn{1}{c|}{$\mathcal{O}(\frac{P}{N})$}   & $\mathcal{O}(\frac{P}{N})$                      \\ \hline
\begin{tabular}[c]{@{}c@{}}Comp. Latency\\ (per sequence)\end{tabular}    & $\mathcal{O}(C)$             & $\mathcal{O}(\frac{C}{N})$   & $\mathcal{O}(\frac{C}{N})$   & \multicolumn{1}{c|}{$\mathcal{O}(C)$}             & $\mathcal{O}(\frac{C}{N})$                      \\ \hline
\begin{tabular}[c]{@{}c@{}}Comm. Volume\\ (per sequence)\end{tabular}     & \textit{None}                & $2LSH$                       & $4LSH$                       & \multicolumn{2}{c}{$(N-1)SH$}                                                                       \\ \toprule[1pt]
\end{tabular}
\vspace{-10pt}
\end{table}

\begin{table}[t!]\setlength{\tabcolsep}{4.1pt}
\centering
\setlength{\abovecaptionskip}{0cm}
\setlength{\belowcaptionskip}{-0.1cm}
\caption{Comm.-to-comp. ratio of various parallelism methods.}
    \label{tab:comm-comp}
\begin{tabular}{ccccccc}
\toprule[1pt]
\multirow{2.5}{*}{\begin{tabular}[c]{@{}c@{}}\textbf{Model}\\ \textbf{Name}\end{tabular}}      & \multirow{2.5}{*}{\begin{tabular}[c]{@{}c@{}}\textbf{Network}\\ \textbf{Bandwidth}\end{tabular}} & \multicolumn{5}{c}{\textbf{Communication-to-Computation Ratio}} \\ \cmidrule(lr){3-7}
                            &                          & SP   & TP   & DT \cite{wei2024communication}   & Galaxy \cite{ye2024galaxy}    & \textbf{\textcolor{redbrown}{Jupiter}}\\ \midrule[0.4pt]
\multirow{2}{*}{Llama2-7B}  & 100Mbps                  & 8.16     & 6.96     & 3.48                & 5.19          &  \textbf{0.08}\\
                            & 1Gbps                 & 0.92     &  0.88    &  0.45               &  0.69         & \textbf{0.01}  \\ \midrule[0.4pt]
\multirow{2}{*}{Llama2-13B} & 100Mbps                  &  5.71    & 6.06     & 3.03                & 4.63          & \textbf{0.05}  \\
                            & 1Gbps                 &  0.73    &  0.81    &  0.38               &  0.56         & \textbf{0.01} \\ \toprule[1pt]
\end{tabular}
\vspace{-15pt}
\end{table}

\begin{figure*}[t!]
    \setlength{\abovecaptionskip}{-0.1cm}
    \centering
    \includegraphics[width=0.92\linewidth]{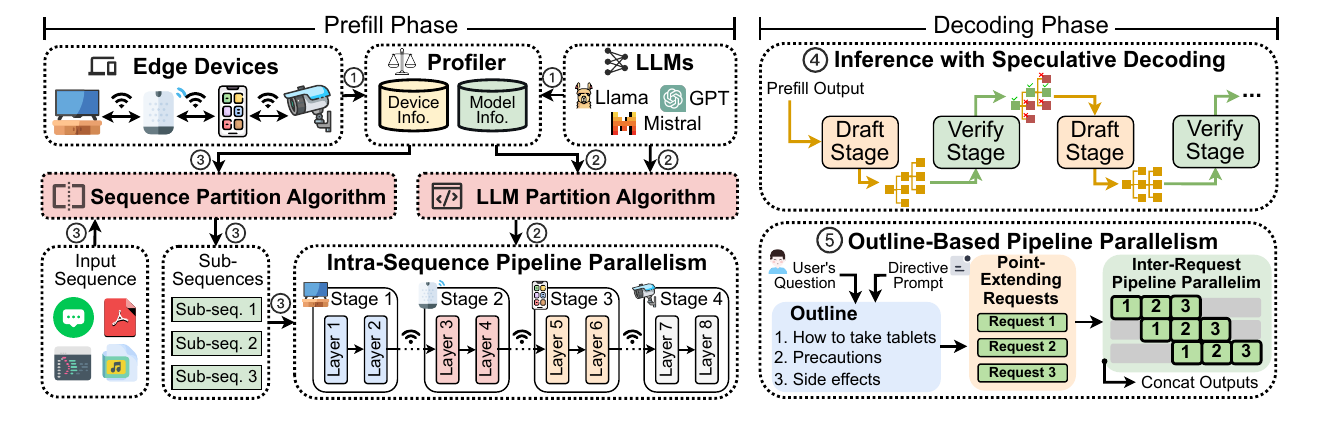}
    \caption{\texttt{Jupiter} system overview.}
    \label{fig:overview}
    \vspace{-15pt}
\end{figure*}

\subsubsection{Issues of Existing Collaborative Edge Inference Systems} 
Most existing collaborative edge inference systems \cite{ye2024galaxy, wei2024communication} employ \textit{TP} and \textit{SP} as the primary principle for parallel LLMs inference.
However, employing \textit{TP} and \textit{SP} involves multiple tensor synchronization in each decoder layer, resulting in significant communication overhead. 
Specifically, Table \ref{tab:comm-comp} summarizes the communication-to-computation latency ratios observed during single-sequence prefilling using various parallelism methods on an edge platform with four Jetson Xavier NX \cite{jetson-NX}. We observe that the ratio can reach up to 8.2 times for \textit{TP} and \textit{SP}-based methods under typical edge network bandwidth.
Despite efforts by these systems to design sophisticated communication optimization techniques, our evaluation in \S \ref{sec:eva} has revealed that communication time still remains a bottleneck under low-bandwidth edge environments.
Besides the aforementioned methods, a few research efforts \cite{zhang2024edgeshard} have explored the use of \textit{PP} to orchestrate edge devices. However, for single-sequence requests, these methods ultimately degrade to serial inference, preventing the concurrent utilization of multiple edge devices.
Moreover, all of the aforementioned works have focused solely on optimizing the prefill phase, neglecting the autoregressive decoding phase, which is also a critical part of the generative LLMs' inference process.

\subsection{Design Goal and Technical Challenges}
As summarized earlier, existing collaborative edge systems for generative LLM inference exhibit limitations. Alternatively, we revisit these limitations and endeavor to propose a more fast, scalable and resource-efficient collaborative edge AI system, guided by the following design goals: 
(1) Support parallel acceleration of single-sequence inference during both the prefill and autoregressive decoding phases.
(2) Achieve scalable memory footprint reduction per device by distributing the LLM parameters across participating devices.
(3) Minimize tensor exchanges between devices to maintain high inference performance even in extremely low-bandwidth edge environments.
To achieve the above design goals, instead \textit{TP}, we adopt a pipelined architecture as a principle to orchestrate collaborative edge devices. Each device holds a portion of the parameters, enabling the collective memory of multiple devices to support the target LLM. Furthermore, pipelined architecture is highly communication-efficient, as devices exchange only a small set of intermediate activations with their neighbors. 
Despite the opportunities, adopting pipelined architecture still suffers from the following non-trivial challenges:

\begin{itemize}[leftmargin=*]
    \item \textit{How to accelerate the prefill phase with pipeline parallelism?}
    The prefill phase is computation-intensive and incurs significant time-to-first-token latency, particularly for long prompts. This motivates us to augment available computing power with collaborative edge devices. However, traditional pipelined inference degrades to serial inference for single-sequence requests, necessitating the exploration of new parallel opportunities for our pipelined architecture.
    \item \textit{How to accelerate the autoregressive decoding phase with pipeline parallelism?} Token-by-token autoregressive decoding is notorious for its prolonged inference process and the inherent difficulty in parallelization. Effectively leveraging edge resources in a pipeline manner poses a significant challenge to the decoding acceleration.
\end{itemize} 

\section{Jupiter System Overview}
Fig. \ref{fig:overview} illustrates an overview of our proposed \texttt{Jupiter} system, which incorporates dedicated designs to optimize both the prefill and decoding phases. 
In the prefill phase, \texttt{Jupiter} profiler first conducts an LLM prefill process using calibration sequences with varying lengths on the edge devices to record the run-time traces necessary for parallelism planning algorithm (\circledwhite{1}). 
LLM partition algorithm takes the target LLM and profiling results as inputs to generate an optimal pipelined partition, considering both the resource heterogeneity and memory budget of edge devices (\circledwhite{2}). 
The input sequence is processed by the sequence partition algorithm, which divides it into multiple sub-sequences. These sub-sequences are then concurrently fed into the inference pipeline, enabling resource-efficient intra-sequence parallel inference (\circledwhite{3}).
\texttt{Jupiter} incorporates two modules designed to maximize the utilization of computational resources by fully exploiting the parallelism potential inherent in the decoding process.
Specifically, we first incorporate speculative decoding into our collaborative edge inference system to enhance resource efficiency by processing multiple candidate tokens in parallel (\circledwhite{4}). 
Next, to further boost parallelism potential by leveraging multiple edge devices concurrently during the decoding phase, we introduce an outline-based pipeline parallel decoding method (\circledwhite{5}).

\section{Parallel Acceleration for Prefill Phase}

\subsection{Intra-Sequence Pipeline Parallelism for Generative LLMs}

\subsubsection{Pipelined Inference for LLMs}
\texttt{Jupiter} adopt a pipelined architecture \cite{zhang2024edgeshard} as a principle to orchestrate collaborative edge devices, which involves partitioning the decoder layers of an LLM into multiple \textit{stages}. Each stage contains a stage model composed of a set of consecutive decoder layers and is mapped to a separate edge device that performs the forward pass (FP) for the corresponding stage model. 
Single-sequence requests are prevalent in edge intelligence services; in this setup, only one device is active at any given time, as depicted in Fig. \ref{fig:pipeline}(\circledwhite{1}).
Ideally, we aim for all edge devices to be active concurrently to fully exploit their resource potential.

\subsubsection{Opportunities of Intra-Sequence Parallel Inference} 
\begin{figure}[t!]
    \setlength{\abovecaptionskip}{-0.1cm}
    \centering
    \includegraphics[width=0.95\linewidth]{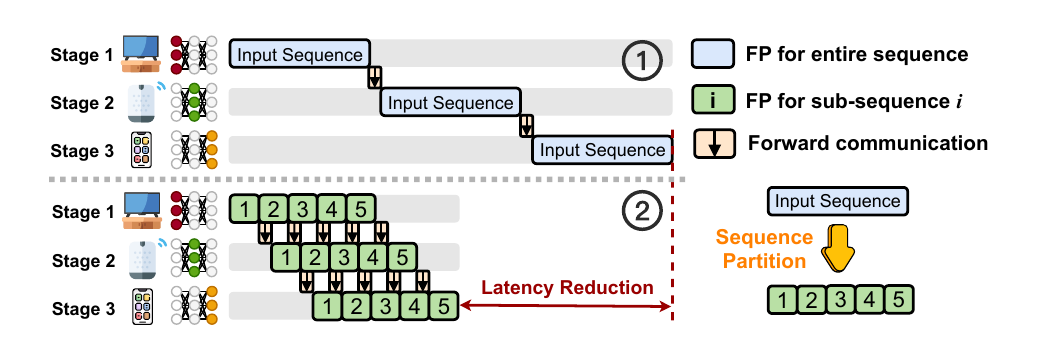}
    \caption{An illustration of pipelined inference with three edge devices.}
    \label{fig:pipeline}
    \vspace{-5pt}
\end{figure}

To fully boost the parallel potential of our pipelined architecture, we propose to partition the input sequence into multiple sub-sequences along the sequence dimension and inject them into the pipeline concurrently to increase parallelism, as depicted in Fig. \ref{fig:pipeline}(\circledwhite{2}). 
Although the computation of the \textit{QKV Project} and \textit{FFN} modules in decoder layers depends on each individual sub-sequence, the \textit{self-attention} operation in the \textit{MHA} module requires calculating dependencies and relationships between tokens across the entire input sequence, making intra-sequence pipeline parallelism a non-trivial task.

\begin{figure}[t!]
    \setlength{\abovecaptionskip}{-0.1cm}
    \centering
    \includegraphics[width=0.95\linewidth]{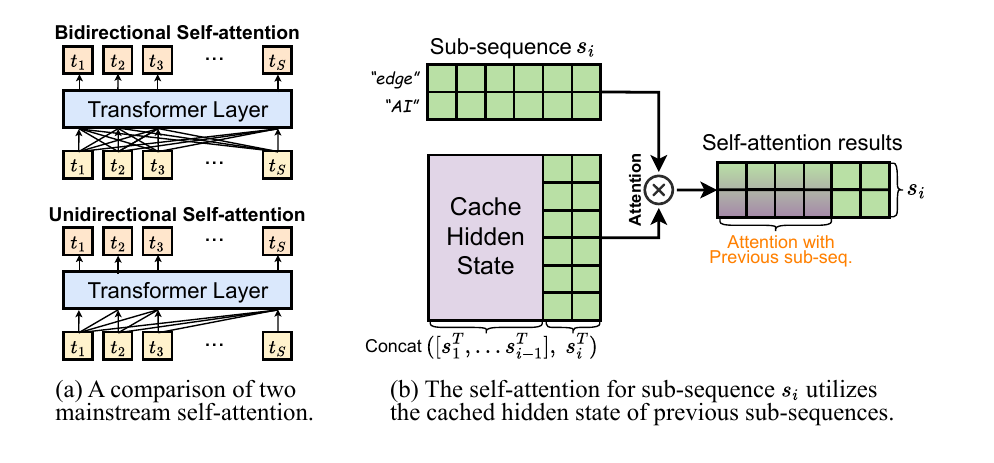}
    \caption{An illustration of opportunities of intra-sequence parallel inference.}
    \label{fig:attention}
    \vspace{-15pt}
\end{figure}

In this work, we leverage the key observation that the causal decoder, the predominant architecture used in current generative LLMs, employs a unidirectional attention mask to ensure that each input token can only attend to previous tokens.
Consequently, for an input sequence $(t_1, t_2,...)$, the computation of self-attention of a token $t_i$ involves only the preceding tokens $t_1, ..., t_{i-1}$, as depicted in Fig. \ref{fig:attention}(a). 
This property brings opportunities for pipeline parallelism within a single input sequence. Specifically, we partition an input sequence into $M$ sub-sequence: $(s_1, s_2,...)$, where each sub-sequence consists of a set of consecutive tokens. 
We then inject all sub-sequences into the pipeline sequentially. During inference, each stage caches the hidden states of every sub-sequence. When computing for $s_i$, we utilize the cached hidden states of $s_1, ..., s_{i-1}$ to ensure correct self-attention results for $s_i$, as depicted in Fig. \ref{fig:attention}(b).

\subsection{Resource-Efficient Parallelism Planning}
In this section, we detail the parallelism planning for optimal partitioning of LLMs and input sequences.

\subsubsection{Selecting Optimal LLMs Partition} 
\label{sec:llm-partition}
To enable pipelined inference, we need to partition the target LLM into multiple stages and map them to edge devices. As with any pipeline, the steady-state throughput is determined by the execution time of the slowest stage. Consequently, we endeavor to partition the LLM into balanced stages. 
We consider an LLM consisting of $L$ layers (embedding, decoder and output head, etc.) and denote $\mathcal{D}$ as an ordered set of all devices involved in parallel inference. $\mathcal{D}_n=\{d_1,...,d_n\}$ denotes the subset of first $n$ device in $\mathcal{D}$. $A(i\rightarrow j, \mathcal{D}_n)$ denote the time taken by the slowest stage in the optimally balanced sub-pipeline between layer $i$ to $j$ with $\mathcal{D}_n$. The goal of our algorithm is to estimate $A(1\rightarrow L, \mathcal{D})$.
To solve this balanced partitioning problem, we break the pipeline into sub-pipelines and leverage the idea of dynamic programming. The formula of the dynamic programming algorithm can be written as:
\vspace{-5pt}
\begin{equation}
    A(1\to y,\mathcal{D}_n)=\min\limits_{1\leqslant l<y}\max\left\{\begin{array}{l}A(1\to l,\mathcal{D}_{n-1}),\\T(l+1\to y,d_n),\end{array}\right.
    \vspace{-2pt}
\end{equation}
where $ T\left(i \rightarrow j, n\right)=\sum_{l=i}^{j} t_{l}^{n}$ represents the time taken by device $d_n$ to process layers $i$ through $j$. $t_l^n$ is averaged from profiling on physical edge devices using a calibration dataset. If the memory required for the stage model spanning layers $i$ through $j$ (including LLM parameters and \textit{KVCache}) exceeds the memory budget of device $d_n$, then $T\left(i \rightarrow j, n\right)=+\infty$.

\subsubsection{Selecting Optimal Sequence Partition}
As previously discussed, we can partition the input sequence into multiple sub-sequences to boost parallelism. However, partitioning the input sequence is not trivial for the following reasons:
(1) Increasing the number of sub-sequences can efficiently boost parallelism and reduce pipeline bubbles. However, this results in shorter sub-sequences, which may underutilize the mobile accelerator (e.g., GPU, NPU), potentially leading to longer processing times.
(2) Partitioning the sequence into sub-sequences of equal length for pipelining is not optimal. As previously discussed, the self-attention of sub-sequence $s_i$ depends on previous $s_1,...,s_{i-1}$. Consequently, later sub-sequences carry a heavier computational load than earlier ones.
(3) The input sequence lengths vary across different requests, necessitating the determination of the optimal partitioning strategy for each length to accommodate varying tasks.

To overcome above challenges, we design a sophisticated sequence partition algorithm to achieve optimal pipeline efficiency. 
We denote the maximum length of the input sequence that the target LLM can process as $S_{max}$. 
To avoid underutilizing devices, we first profile the accelerator utilization on each device at various input sequence lengths ($<S_{max}$). We then decide the minimum sub-sequence length, denoted as $b$, that prevents underutilization of all mobile accelerators.

\begin{figure}[t!]
    \setlength{\abovecaptionskip}{-0.1cm}
    \centering
    \includegraphics[width=0.95\linewidth]{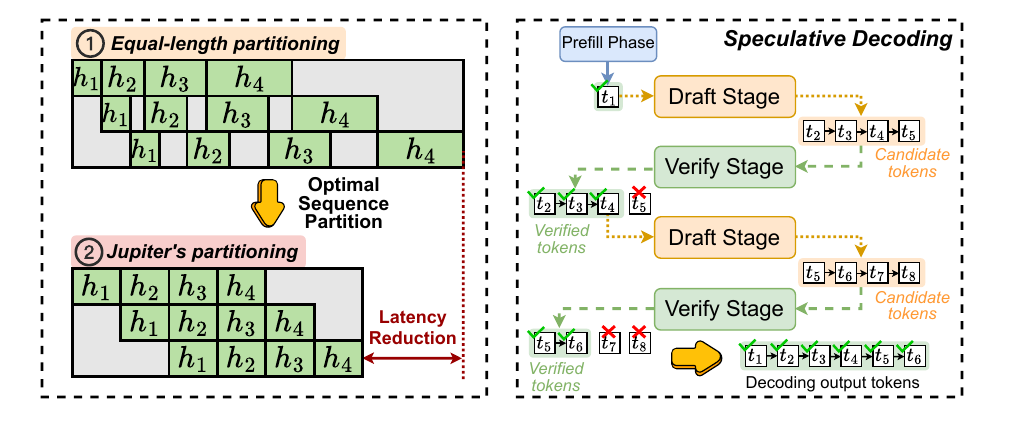}
    \caption{Left: Comparison between equal-length and \texttt{Jupiter}'s partition. Right: An illustration of the decoding phase with Speculative Decoding.}
    \label{fig:par_spec}
    \vspace{-5pt}
\end{figure}

After optimal LLM partitioning, we obtain a partitioned LLM with multiple stage models, each mapped to an edge device and having exact same computational latency.
We use $h_i$ to represent the inference latency of sub-sequence $s_i$ at each stage, as shown in Fig. \ref{fig:par_spec}(Left).
As detailed earlier, in our intra-sequence parallel inference, the inference latency of each sub-sequence $s_i$ depends on $s_1, ..., s_{i-1}$ and itself. 
We use $q(x,y)$ to denote the inference latency for a sub-sequence of length $x$, given the total length $y$ of its previous sub-sequences. Thus, $h_i$ can be expressed as $h_i = q(|s_i|, \sum_{j=1}^{i-1}|s_j|)$.
We meticulously profile $q(x,y)$ on a realistic edge platform under various sequence lengths $x$ and $y$. The profiling overhead can be linearly reduced by conducting concurrent profiling on multiple devices and approximating results through interpolation.

The goal of our algorithm is to find optimal partition schemes for sequences of varying lengths ($<S_{max}$), ensuring that the inference latency of each sub-sequence is as balanced as possible while each sub-sequence length exceeds $b$ to avoid device underutilization.
We leverage dynamic programming to achieve the goal.
We denote $W(i\rightarrow,j, k)$ as the inference latency of the slowest sub-sequence in the optimal partitioning of the sequence from token $i$ to $j$ into $k$ sub-sequences.
The formula of the dynamic programming can be written as:
\vspace{-5pt}
\begin{equation}
    W(1\to y, k)=\min\limits_{1\leqslant l<y}\max\left\{\begin{array}{l}W(1\to l, k-1),\\T^*(y-l, l).\end{array}\right.
    \label{eq:dp-seq}
\end{equation}
\begin{equation}
T^*\left(y-l,l\right)=\left\{
\begin{array}{ll}
+\infty, \quad \text{if} \ y-l<b, \\
q(y-l, l), \quad \text{else}.
\end{array}
\right.
\vspace{-2pt}
\end{equation}
When solving for Eq. \ref{eq:dp-seq}, the sentence length $y$ is iterated from $1$ to $S_{max}$, and $k$ is iterated from $1$ to $4|\mathcal{D}|$. We set the maximum number of sub-sequences to $4|\mathcal{D}|$, which effectively boosts the pipeline parallelism while preventing excessive planning algorithm runtime.
For each sentence length, we record the optimal balanced partitioning strategy for dividing the sentence into a varying number of sub-sequences.

Upon the completion of dynamic programming process, we need to select the optimal number of sub-sequences $k$ for each input length $y$. We observe from Fig. \ref{fig:par_spec}(Left) that the total inference latency for a sequence of length $y$ partitioned into $k$ sub-sequences can be estimated by:
\vspace{-5pt}
\begin{equation}
     \text{Latency} = \sum_{i=1}^{k}h_i+ (|\mathcal{D}|-1)\times W(1\rightarrow y,k).
     \label{equ:choose_k}
     \vspace{-2pt}
\end{equation}
We choose $k$ to minimize Eq. \ref{equ:choose_k} for each sequence length $y$.

\begin{figure}[t]
    \setlength{\abovecaptionskip}{-0.1cm}
    \centering
    \includegraphics[width=0.95\linewidth]{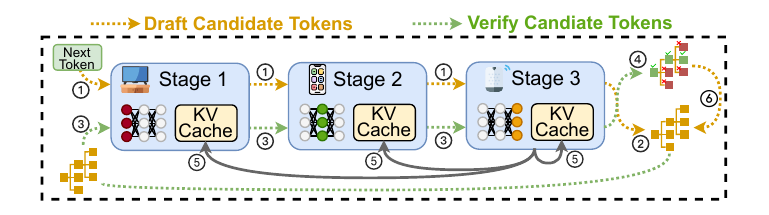}
    \caption{A workflow of our collaborative inference with speculative decoding.}
    \label{fig:workflow}
    \vspace{-15pt}
\end{figure}

\subsubsection{Complexity} 
The time complexity for our optimal LLM partition is $\mathcal{O}(L^2|\mathcal{D}|)$, while for the optimal sequence partition it is $\mathcal{O}(S_{max}^2|\mathcal{D}|)$. In our experiments, the entire planning process is completed in under five minutes on an edge device. Notably, parallelism planning is a one-shot offline process and its outputs can be stored and reused. The overhead of it can be amortized across thousands of inference iterations.

\section{Collaborative Inference for Decoding Phase}
The decoding latency of LLMs has become a substantial obstacle to high-quality human-computer interactions. This latency stems from the token-by-token generation required by autoregressive decoding, causing significant delays in producing long outputs.
To accelerate LLM decoding, an intuitive way involves leveraging idle computational resources to enhance parallelism. \texttt{Jupiter} introduces two parallel decoding strategies to accelerate the decoding phase.

\subsection{Collaborative Inference with Speculative Decoding}
Speculative decoding \cite{xia2024unlocking, stern2018blockwise, leviathan2023fast, chen2023accelerating, cai2024medusa} has emerged as a promising paradigm for accelerating decoding. In each decoding step, speculative decoding first drafts multiple candidate tokens efficiently, speculating on future decoding steps. This is achieved using extra lightweight heads atop current LLM backbone (\textit{Self-Drafting}) or a small independent draft model (\textit{Independent Drafting}). These candidate tokens are then verified in parallel by original LLM to ensure the outputs align with the original distribution, as illustrated in Fig. \ref{fig:par_spec}(Right).
By leveraging parallel token generation, speculative decoding significantly reduces the total number of decoding steps required.

\begin{figure}[t!]
    \setlength{\abovecaptionskip}{-0.1cm}
    \centering
    \includegraphics[width=0.95\linewidth]{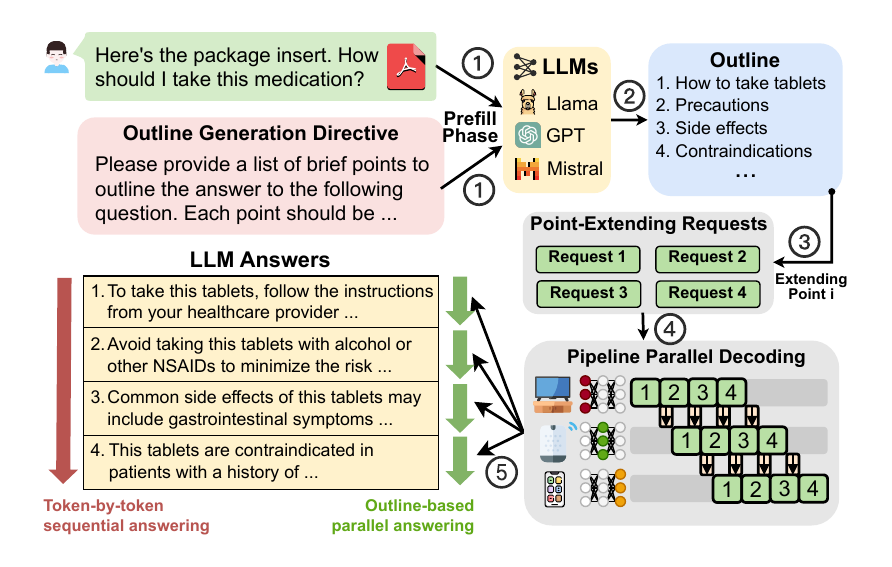}
    \caption{An illustration of our outline-based pipeline parallel decoding.}
    \label{fig:outline}
    \vspace{-15pt}
\end{figure}

\texttt{Jupiter} incorporates the idea of \textit{Self-Drafting} from speculative decoding into our collaborative edge inference system to enhance resource efficiency during decoding phase. Fig. \ref{fig:workflow} illustrated our workflow. Specifically, the next token produced during the prefill phase is fed into the LLM for a forward pass to obtain logits (\circledwhite{1}). The logits will be processed by the FFN heads incorporated atop the LLM to generate multiple candidate tokens in parallel (\circledwhite{2}). The candidate tokens will be transferred from the final stage back to the initial stage and fed into the LLM for a forward pass. The intermediate results of all candidate tokens will be precisely stored in the \textit{KVCache} (\circledwhite{3}).
The posterior probability of each speculative candidate will be evaluated to determine whether it should be accepted or rejected (\circledwhite{4}). Subsequently, the final stage will inform all stages of the rejected candidate tokens, directing them to remove these tokens from the \textit{KVCache} (\circledwhite{5}). 
We extract the logits of the accepted tokens from the output produced in step \circledwhite{3}. These logits are processed by the draft heads to generate a new batch of candidate tokens for the next decoding iteration (\circledwhite{6}).
Our workflow design and implementation minimize redundant LLM calls, ensuring that each draft and verification process requires only one LLM inference. \texttt{Jupiter} supports the flexible plug-and-play replacement of various self-drafting speculative decoding algorithms.
In our evaluation, we incorporate the \textit{token tree}-based speculative decoding algorithm proposed by Medusa \cite{cai2024medusa}.

\subsection{Outline-Based Pipeline Parallel Decoding}
Incorporating speculative decoding with collaborative inference enables parallel processing of multiple candidate tokens, thus attaining improved device utilization.
However, merely applying speculative decoding on pipelined inference yet operates serially, failing to leverage multiple devices concurrently.

To fully exploit idle computational resources at the edge, we intend to further explore the parallel potential during the decoding phase.
Towards that, we borrow wisdom from human thinking, which usually organizes thoughts against questions first and then responds point-by-point.
This routine, in many situations, is more common and efficient than purely sequential answering and has been experimentally verified by many recent explorations \cite{wei2022chain, yao2024tree, long2023large}.
These works typically guide LLMs in generating an explicit chain or tree-like thought process, subsequently eliciting high-quality answers.

Inspired by them, we introduce an outline-based pipeline parallel decoding method, with its workflow illustrated in Fig. \ref{fig:outline}. Specifically, we first concatenate a crafted outline generation directives with the user's question for prefilling (\circledwhite{1}) to guide the LLM in organizing its thoughts and generating an outline of the answers (\circledwhite{2}).
The prefill for static guide prompts can be performed offline in advance and cached into the \textit{KVCache}.
Next, we package each point from the outline into separate point-extending requests. Each request guides the LLM to expand solely on that specific point (\circledwhite{3}).
These requests are then injected into the collaborative inference pipeline concurrently for efficient pipeline parallelism (\circledwhite{4}). 
The \textit{KVCache} of input sequence generated during the prefill phase will be shared across all requests, thereby eliminating redundant computations.
Finally, after all point-extending requests are finished, we will concatenate the outputs from each request to obtain the final answer (\circledwhite{5}).

Note that different tasks (e.g., document summarization, and question answering) can utilize distinct outline generation directives in generating outlines appropriate for parallel inference.
For tasks requiring step-by-step reasoning with chained logical dependencies, such as math and coding, or tasks needing only short answers, the outline-based parallel decoding method may not generate high-quality responses.
Therefore, our system design includes outline-based parallel decoding as a flexible, pluggable module. For problem types less suited to it, the system can automatically decide or let the user choose whether to use it, thus avoiding unsatisfactory results.
In these cases, our inference system automatically defaults to speculative decoding for sequential answering, and experimental evaluations demonstrate that Jupiter can still achieve outstanding performance in latency reduction.

\section{Implementation and Evaluation}
\label{sec:eva}

\subsection{Experimental Setups}

\subsubsection{Models and Datasets} 
We evaluate \texttt{Jupiter} using 2 LLMs from the Llama2 series \cite{touvron2023llama}, specifically Llama2-7B and Llama2-13B (both with INT4 quantization). We employ 3 recent assistant-style datasets: LiMA \cite{zhou2024lima}, Vicuna-80 \cite{chiang2023vicuna}, and WizardLM \cite{xu2023wizardlm}. LiMA is used to evaluate inference latency, and all datasets are utilized for assessing generation quality.

\begin{table}[t!]\setlength{\tabcolsep}{4pt}
\centering
\setlength{\abovecaptionskip}{0cm}
    \setlength{\belowcaptionskip}{-0.1cm}
\caption{Specifications of edge devices in experiments.}
\begin{tabular}{cccc}
\toprule[1pt]
\textbf{Edge Device} & \textbf{GPU Processor}     & \textbf{\begin{tabular}[c]{@{}c@{}}Memory\end{tabular}} & \textbf{Power} \\ \hline
Jetson Xavier NX \cite{jetson-NX}           & 384-core NVIDIA Volta  & 8GB & 20W                                                         \\ 
Jetson TX2 \cite{jetson-TX2}         & 256-core NVIDIA Pascal & 8GB & 20W                                                              \\ 
Jetson Nano \cite{jetson-nano}         & 128-core NVIDIA Maxwell   & 8GB  & 10W
                                                               \\ \toprule[1pt]
\end{tabular}
\label{tab:hardware}
\vspace{-15pt}
\end{table}

\begin{table*}[t!]\setlength{\tabcolsep}{4.8pt}
    \centering
    \setlength{\abovecaptionskip}{0cm}
    \setlength{\belowcaptionskip}{-0.1cm}
    \caption{End-to-end generation latency (in seconds) for LiMA dataset including the prefill and decoding phases under various settings.}
    \label{tab:e2e}
    \begin{tabular}{cccccccc|cccccc}
        \toprule[1pt]
        \multirow{2.5}{*}{\makecell{\textbf{Edge} \\ \textbf{Environment}}}
        & \multirow{2.5}{*}{\makecell{\textbf{Network} \\ \textbf{Bandwidth}}}
        & \multicolumn{6}{c}{\textbf{Llama2-7B}}
        & \multicolumn{6}{c}{\textbf{Llama2-13B}}
        \\ \cmidrule(lr){3-8}\cmidrule(lr){9-14}

        &
        & SP
        & M-LM
        & DT
        & Galaxy
        & EdgeShard
        & \textcolor{redbrown}{\textbf{\texttt{Jupiter}}}
        & SP
        & M-LM
        & DT
        & Galaxy
        & EdgeShard
        & \textcolor{redbrown}{\textbf{\texttt{Jupiter}}}
        \\
        \midrule[0.8pt]
        \multirow{4}{*}{\makecell{Homo. \\ Env. A}}
        & 100Mbps
        & 53.5
        & 431.2
        &228.5 
        &427.6 
        & 42.2
        & \textbf{16.5}
        & OOM
        & 503.4
        & 270.1 
        & 496.5
        & 66.2
        & \textbf{26.3}
        \\
        \cmidrule(lr){2-14}
        & 500Mbps
        & 37.4
        & 106.9
        & 66.4
        & 103.9
        & 39.0
        & \textbf{15.2}
        & OOM
        & 130.1
        & 83.4 
        & 125.0 
        & 63.4
        & \textbf{25.2}
        \\
        \cmidrule(lr){2-14}
        & 1Gbps
        & 35.4
        & 66.4
        & 46.1
        & 65.0
        & 38.6
        & \textbf{14.9}
        &  OOM
        & 83.4
        & 60.1
        & 81.3
        & 63.1
        & \textbf{24.9}
        \\
        \midrule[0.8pt]
        \multirow{4}{*}{\makecell{Hetero. \\ Env. B}}
        & 100Mbps
        & 63.1
        & 491.2
        & 288.6
        & 458.3
        & 59.3
        & \textbf{22.4}
        & OOM
        & 624.5
        & 391.2 
        & 566.4
        & 102.4
        & \textbf{38.8}
        \\
        \cmidrule(lr){2-14}
        & 500Mbps
        & 47.0
        & 167.0 
        & 126.4
        & 142.9 
        & 56.1 
        & \textbf{21.4}
        & OOM
        & 251.2
        & 204.5
        & 208.0
        & 99.7 
        & \textbf{37.3}
        \\
        \cmidrule(lr){2-14}
        & 1Gbps
        & 44.8
        & 126.4
        & 106.2
        & 104.9
        & 55.7 
        & \textbf{20.9}
        & OOM
        & 204.5
        & 181.2
        & 165.7
        & 98.3
        & \textbf{36.8}
        \\
        \bottomrule[1pt]
    \end{tabular}
    \vspace{-15pt}
\end{table*}

\subsubsection{Edge Environment Setup}
We use three heterogeneous off-the-shelf edge devices, as listed in Table \ref{tab:hardware}, in our experiments. We evaluate \texttt{Jupiter}'s performance in two realistic edge environments, incorporating both homogeneous and heterogeneous configurations. \textit{Homogeneous Environment A} consists of $4\times$NX, while \textit{Heterogeneous Environment B} comprises $1\times$NX, $2\times$TX2, and $1\times$Nano. We adjust the device-to-device communication bandwidth to simulate the diverse network conditions within realistic edge environments.

\subsubsection{Baseline Methods}
We compare \texttt{Jupiter} with five state-of-the-art parallel LLMs inference method:
\begin{itemize}[leftmargin=*]
    \item \textit{Sequence Parallelism (SP)} \cite{li2023sequence} is pioneering work that proposes \textit{SP} for distributed LLM execution in datacenters.
    \item \textit{Megatron-LM (M-LM)} \cite{narayanan2021efficient} is pioneering work that proposes \textit{TP} for distributed LLM execution in datacenters.
    \item \textit{DeTransformer (DT)} \cite{wei2024communication} is a \textit{TP}-based collaborative edge inference system that strikes a trade-off between communication overhead and inference accuracy by reducing the frequency of tensor synchronization. In our evaluation, we selected the decoupled layers to be half of the total layers.
    \item \textit{Galaxy \cite{ye2024galaxy}} is a collaborative edge inference system that employs \textit{TP} across MHA and FFN modules, with SP applied to the connecting operations. It employs fine-grained overlapping of comm. and comp. to mitigate inference latency.
    \item \textit{EdgeShard \cite{zhang2024edgeshard}} is a collaborative edge inference system that employs pipelined architecture to orchestrate edge devices.
\end{itemize} 
Given that existing inference systems do not optimize the decoding phase, we equipped them with the \textit{naive} token-by-token sequence generation method. During the decoding phase, \textit{SP} will degrade to single-device inference due to the sequence length being one.

\subsection{End-to-End Performance}
\label{sec:e2e-per}
Table \ref{tab:e2e} summarizes the end-to-end generation latency of \texttt{Jupiter} and baselines. We sample prompts from LiMA dataset with an average sequence length of 260 tokens and set the maximum generation length to 64 tokens. The results indicate that \texttt{Jupiter} consistently outperforms all baselines across various models, edge environments, and edge network bandwidths. Specifically, when compared to \textit{TP}-based parallel inference methods like M-LM, DT and Galaxy, \texttt{Jupiter} achieves up to $26.1\times$ latency reduction. 
When compared to \textit{SP}-based parallel inference method like \textit{SP}. \texttt{Jupiter} achieve up to $3.3\times$ latency reduction. \textit{SP} degrades to single-device inference during the decoding phase, avoiding significant communication overhead but wasting resources on idle devices. Additionally, \textit{SP} requires each device to accommodate all parameters, causing out-of-memory (OOM) issues with a 13B model. When compared to pipelined methods like EdgeShard, \texttt{Jupiter} achieves up to a $2.7\times$ reduction in latency by fully leveraging the computational resources of multiple devices concurrently. 
\texttt{Jupiter}'s resource-efficient parallelism planning accounts for the computational resources of heterogeneous devices, consistently outperforming baselines in heterogeneous environment B and achieving a $2.6\times$ to $21.9\times$ latency reduction.

\begin{figure}[t!]
    \setlength{\abovecaptionskip}{-0.1cm}
    \centering
    \includegraphics[width=0.95\linewidth]{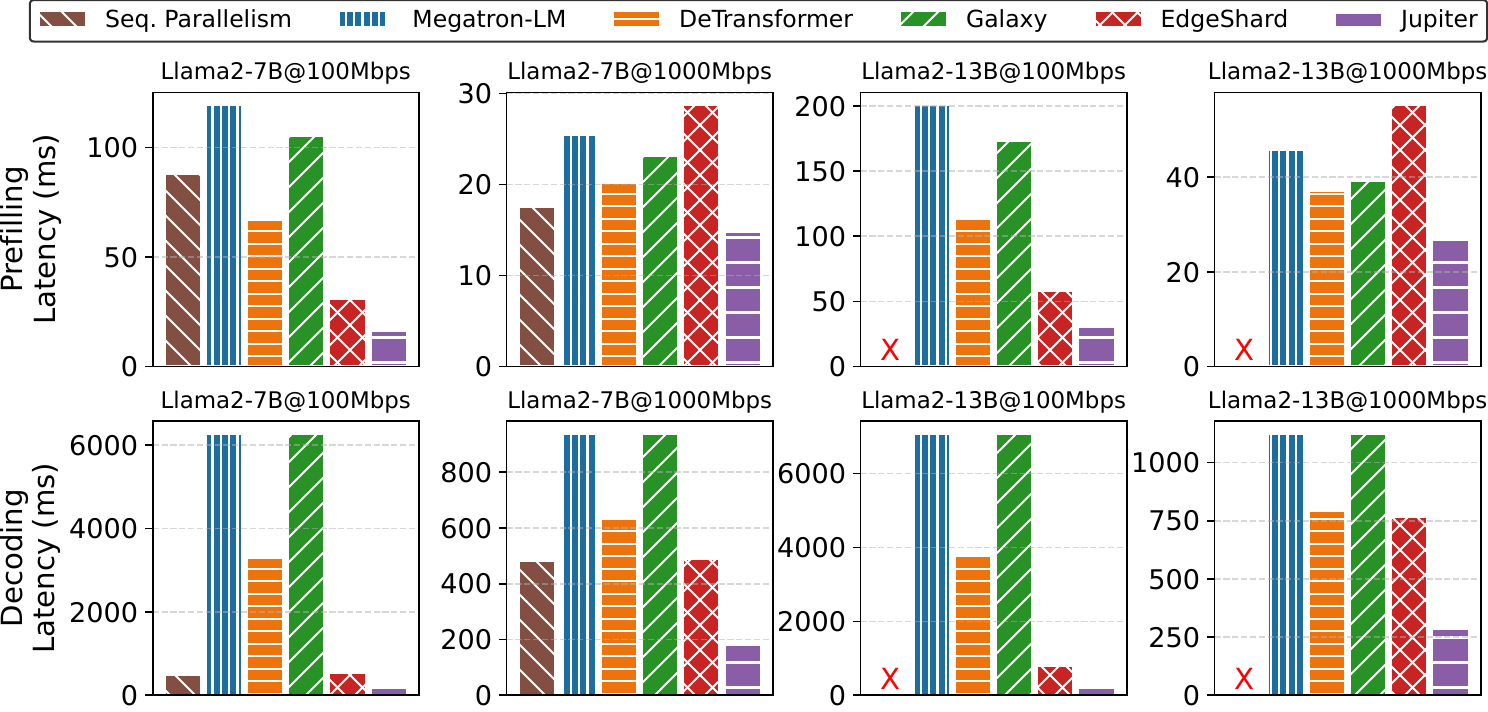}
    \caption{Evaluate in Homogeneous Environment A. The average per-token processing/generation latency in prefill/decoding phase. $\times$ indicates OOM.}
    \label{fig:phase-wise-homo}
\end{figure}

\subsection{Phase-Wise Analysis}
\begin{figure}[t!]
    \setlength{\abovecaptionskip}{-0.1cm}
    \centering
    \includegraphics[width=0.95\linewidth]{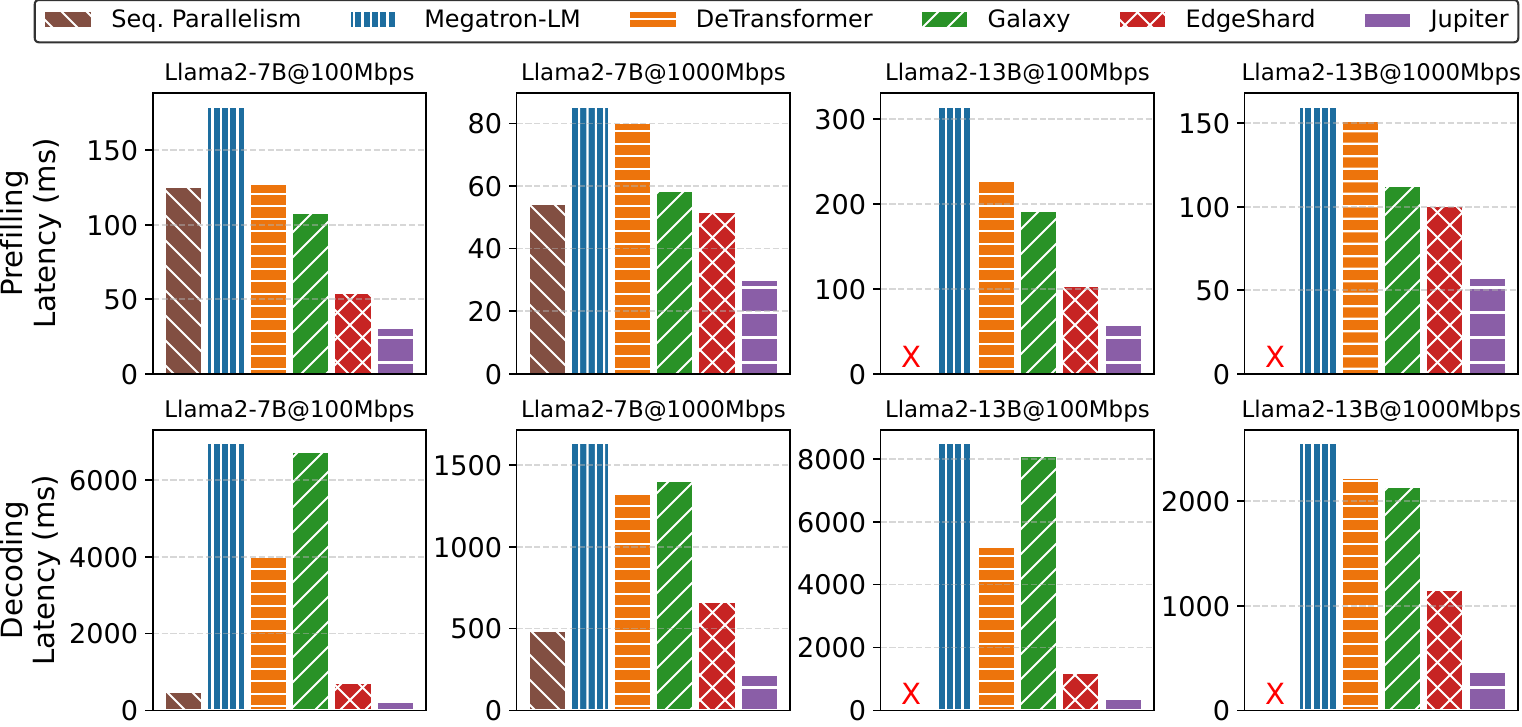}
    \caption{Evaluate in Heterogeneous Environment B. The average per-token processing/generation latency in prefill/decoding phase. $\times$ indicates OOM.
    }
    \label{fig:phase-wise-hetero}
\end{figure}

We further investigate the performance improvements in the prefill and decoding phases separately. In Fig. \ref{fig:phase-wise-homo} and \ref{fig:phase-wise-hetero} we report the average per-token processing latency during the prefill phase and generation latency during the decoding phase in homogeneous and heterogeneous environments, respectively. 
For the prefill phase, \texttt{Jupiter} achieves a $1.4\times$ to $7.4\times$ reduction in latency compared to the baselines. 
Despite efforts by state-of-the-art edge inference systems like DT and Galaxy to design sophisticated communication optimization techniques, such as fine-grained communication-computation overlapping, these methods still perform poorly in bandwidth-constrained edge networks.
For the decoding phase, \texttt{Jupiter} significantly outperforms the baselines, achieving a $2.9\times$ to $33.2\times$ reduction in latency.
In TP and SP-based parallel architectures, the high communication-to-computation ratio is further exacerbated during token-by-token autoregressive generation, while pipelined architectures fail to concurrently utilize the computational resources of multiple devices. These issues collectively amplify the severe resource wastage problem during decoding phase. 
In contrast, \texttt{Jupiter}'s system design fully boosts parallelism potential during autoregressive generation, significantly accelerating the decoding phase.

\subsection{Scalability}
\begin{figure}[t!]
    \setlength{\abovecaptionskip}{-0.1cm}
    \centering
    \includegraphics[width=\linewidth]{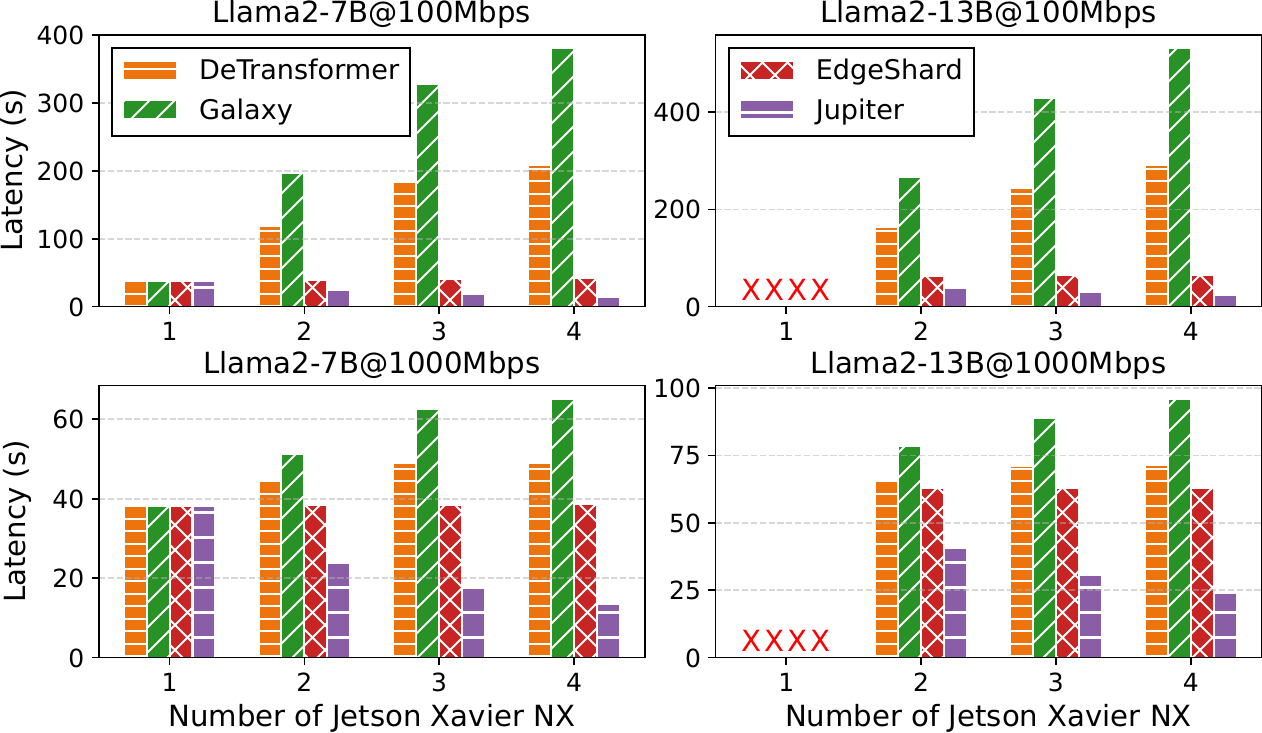}
    \caption{End-to-end inference latency with a varing number of Jetson Xavier NX under 100Mbps and 1Gbps network bandwidths. $\times$ indicates OOM.}
    \label{fig:scalability}
\end{figure}

We analyze the scalability of \textit{Jupiter} on a 4-node homogeneous Jetson Xavier NX cluster. We use the same set of input prompts as in \S \ref{sec:e2e-per}, with a maximum generation length of 64 tokens. The results are summarized in Fig. \ref{fig:scalability}. We observe that \texttt{Jupiter} exhibits substantial scalability even under a bandwidth-limited (100Mbps) edge environment. When compared to existing state-of-the-art collaborative edge inference frameworks, \texttt{Jupiter} can achieve up to $23.7\times$ latency reduction. 
The high communication-to-computation ratio of these frameworks makes it challenging to scale resource-efficiently in bandwidth-constrained edge environments. 
The scalability analysis indicates that our \texttt{Jupiter} framework enables the addition of more edge devices to aggregate computational resources, allowing for parallel acceleration of inference and leveraging collective memory to support larger LLMs.

\subsection{Decoding Speedup and Generation Quality Assessment}
\subsubsection{Decoding Speedup Analysis} 
We investigate the performance boost of each decoding optimization module on homogeneous edge environment A. In our evaluation, we incorporate the \textit{token tree}-based speculative decoding algorithm proposed by Medusa \cite{cai2024medusa}, which utilizes five draft heads with top-1 prediction. We conduct an ablation study to assess the contributions of speculative decoding and outline-based pipeline parallelism, as depicted in Table \ref{tab:speedup-naive}.
We observe that both modules achieve effective decoding acceleration, with an overall speedup ratio of up to $3.9\times$.

\subsubsection{Generation Quality Assessment} Previous works \cite{leviathan2023fast, chen2023accelerating, cai2024medusa} have shown that speculative decoding achieves nearly lossless generation results compared to naive token-by-token sequence generation, as its verification process will correct the output distribution. 
In this section, we separately assess the generation quality of our outline-based pipeline parallel decoding method. 
We select the widely adopted LLM-based evaluation framework FastChat \cite{zheng2024judging} to compare the answer quality of naive token-by-token sequence generation (\textit{naive generation}) and our outline-based parallel generation. FastChat, empowered by GPT-4o, will assign a quality score between 1 and 10 for each answer.
As summarized in Table \ref{tab:quality1}, we compared the generation quality of naive generation and our method on the Vicuna-80, WizardLM, and LiMA datasets. 
We observe that our outline-based pipeline parallel decoding significantly reduces latency while maintaining comparable generation quality to that of naive generation.
However, across all three datasets, the overall generation quality of our outline-based parallel decoding method is slightly lower than that of naive generation.
Therefore, we conducted further experiments on the Vicuna-80 dataset to analyze the reasons for the lower quality. We manually selected five question categories from the Vicuna-80 dataset: \textit{Generic}, \textit{Knowledge}, \textit{Counterfactual}, \textit{Coding}, and \textit{Math}, and evaluated them using FastChat, as shown in Table \ref{tab:quality2}. Our results show that for \textit{Generic}, \textit{Knowledge}, \textit{Counterfactual} questions, our outline-based method achieved comparable or superior generation quality. However, for tasks requiring step-by-step reasoning with chained logical dependencies, such as \textit{Coding} and \textit{Math}, our outline-based parallel decoding exhibited lower generation quality.
Therefore, our system design incorporates outline-based parallel decoding as a flexible, pluggable module. For problem types less suited to it, the system can automatically decide or let the user choose whether to use it, thus avoiding unsatisfactory answers.

\section{Related Work}
\subsection{Collaborative Edge Computing for DNN Inference}
CoEdge and DeepThings \cite{zeng2020coedge, zhao2018deepthings} enable the distributed execution of CNN-based inference applications on resource-constrained edge clusters. Galaxy and DeTransformer \cite{ye2024galaxy, wei2024communication, du2024co} utilize \textit{TP} to accelerate transformer inference with collaborative edge devices. EdgeShard \cite{zhang2024edgeshard} orchestrates edge devices in a pipelined manner for sequential inference.

\subsection{Parallel LLMs Execution Architectures}
\textit{DP} \cite{li2014communication, rajbhandari2020zero} is the most extensively used distributed execution method in datacenter. Gpipe \cite{huang2019gpipe} and subsequent works \cite{narayanan2019pipedream, fan2021dapple} leverage \textit{PP} to address the memory challenges associated with executing LLMs that contain billions of parameters. Megatron \cite{narayanan2021efficient} initially introduced \textit{TP} as a principle to parallelize the execution of the transformer-based LLMs. 
To overcome the limitations of handling long sequences,  \cite{li2023sequence, li2021terapipe, ma2024hpipe} borrow the concept of intra-sequence parallelism from recurrent neural networks, splitting long sequences across multiple devices for concurrent execution.

\begin{table}[t!]\setlength{\tabcolsep}{4.8pt}
\setlength{\abovecaptionskip}{-0.1cm}
 \centering
 \caption{Speedup over naive sequential generation. SD: Speculative Decoding. OP: Outline-based parallel decoding.
}
\label{tab:speedup-naive}
\begin{tabular}{ccccc}
\toprule[1.2pt]
\multirow{2.5}{*}{\textbf{Model}} & \multicolumn{4}{c}{\textbf{Speedup Over Naive}}                      \\ \cmidrule(lr){2-5}
                                                                      & Naive & Jupiter w/o OP & Jupiter w/o SD & \textbf{\textcolor{redbrown}{Jupiter}} \\ \toprule[0.7pt]
Llama2-7B                                                             & $1.0\times$      &  $1.8\times$       & $2.3\times$             &  $3.6\times$               \\ 
Llama2-13B                                                            & $1.0\times$      &  $2.0\times$       & $2.4\times$             &  $3.9\times$              \\ \toprule[1.2pt]
\end{tabular}
\end{table}

\begin{table}[t!]
 \setlength{\abovecaptionskip}{-0.1cm}
 \centering
 \caption{Overall Answers Quality of Naive and Jupiter's Method.
}
    \includegraphics[width=\linewidth]{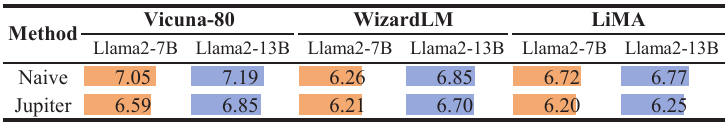}
\label{tab:quality1}
\end{table}

\begin{table}[t!]
 \setlength{\abovecaptionskip}{-0.1cm}
 \centering
 \caption{Answers quality on various question categories in Vicuna-80.
}
    \includegraphics[width=\linewidth]{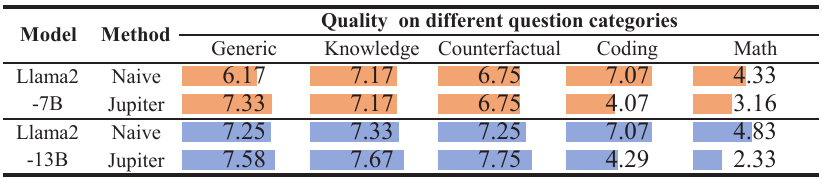}
\label{tab:quality2}
\end{table}

\subsection{Speculative Decoding for LLMs Inference Acceleration} 
Blockwise decoding \cite{stern2018blockwise} is a pioneering work proposing the \textit{Draft-then-Verify} paradigm.
\cite{leviathan2023fast, chen2023accelerating, xia2023speculative} further unleashes its potential and utilizes independent lightweight LLMs to perform the drafting task both accurately and efficiently. While leveraging an external drafter model offers considerable advantages, obtaining an appropriate draft model remains challenging. To address that, numerous studies suggest leveraging the target LLM itself for efficient self-drafting \cite{cai2024medusa, miao2024specinfer, santilli2023accelerating}.

\section{Conclusion}
This paper introduces \texttt{Jupiter}, a fast and scalable collaborative edge inference framework for generative LLMs. \texttt{Jupiter} employs a communication-efficient and resource-scalable pipelined architecture, combined with sophisticated system design, to parallelize and accelerate the prefill and decoding phases. Our extensive evaluation demonstrates that \texttt{Jupiter} achieves up to $26.1\times$ end-to-end generation latency reduction compared to state-of-the-art methods.

\balance
\newpage

\bibliographystyle{IEEEtran}
\normalem
\bibliography{reference}

\end{document}